\documentclass[prd,aps,reprint,superscriptaddress]{revtex4-1}

\usepackage[utf8]{inputenc}
\usepackage{amsmath}    
\usepackage{graphicx}   
\usepackage{verbatim}   
\usepackage{color}      
\usepackage{subfigure}  
\usepackage{hyperref}   
\usepackage{lipsum}
\usepackage{physics}
\usepackage{siunitx}

\usepackage{lineno}

\begin{document}
\newcommand{\authorlist}{
\author{N. Agafonova}
\affiliation{INR - Institute for Nuclear Research of the Russian Academy of Sciences, RUS-117312 Moscow, Russia}
\author{A. Alexandrov}
\affiliation{INFN Sezione di Napoli, I-80126 Napoli, Italy}
\author{A. Anokhina}
\affiliation{SINP MSU - Skobeltsyn Institute of Nuclear Physics, Lomonosov Moscow State University, RUS-119991 Moscow, Russia}
\author{S. Aoki}
\affiliation{Kobe University, J-657-8501 Kobe, Japan}
\author{A. Ariga}
\affiliation{Albert Einstein Center for Fundamental Physics, Laboratory for High Energy Physics (LHEP), University of Bern, CH-3012 Bern, Switzerland}
\author{T. Ariga}
\affiliation{Albert Einstein Center for Fundamental Physics, Laboratory for High Energy Physics (LHEP), University of Bern, CH-3012 Bern, Switzerland}
\affiliation{Faculty of Arts and Science, Kyushu University, J-819-0395 Fukuoka, Japan}
\author{A. Bertolin}
\affiliation{INFN Sezione di Padova, I-35131 Padova, Italy}
\author{C. Bozza}
\affiliation{Dipartimento di Fisica dell'Universit\`a di Salerno and ``Gruppo Collegato''  INFN, I-84084 Fisciano (Salerno), Italy}
\author{R. Brugnera}
\affiliation{INFN Sezione di Padova, I-35131 Padova, Italy}
\affiliation{Dipartimento di Fisica e Astronomia dell'Universit\`a di Padova, I-35131 Padova, Italy}
\author{S. Buontempo}
\affiliation{INFN Sezione di Napoli, I-80126 Napoli, Italy}
% ...
}

\title{Final results on neutrino oscillation parameters from the OPERA experiment in the CNGS beam}

\authorlist
\collaboration{OPERA Collaboration}

\begin{abstract}
The OPERA experiment has conclusively observed the appearance of tau neutrinos in the muon neutrino CNGS beam. Exploiting the OPERA detector capabilities, it was possible to isolate high purity samples of $\nu_{e}$, $\nu_{\mu}$ and $\nu_{\tau}$ charged current weak neutrino interactions, as well as neutral current weak interactions. In this Letter, the full dataset is used for the first time to test the three-flavor neutrino oscillation model and to derive constraints on the existence of a light sterile neutrino within the framework of the $3+1$ neutrino model. For the first time, tau and electron neutrino appearance channels are jointly used to test the sterile neutrino hypothesis. A significant fraction of the sterile neutrino parameter space allowed by LSND and MiniBooNE experiments is excluded at 90\% C.L. In particular, the best-fit values obtained by MiniBooNE combining neutrino and antineutrino data are excluded at 3.3 $\sigma$ significance. 
\end{abstract}

\maketitle

\section{Introduction}
The OPERA experiment \cite{Acquafredda:2009zz} was designed to conclusively observe the appearance of tau neutrino in the high purity muon neutrino CNGS beam \cite{CNGSbeam:1998, Baldy:1999dc}, in the parameter region indicated by Super-Kamiokande \cite{Fukuda:1998mi} and MACRO \cite{Ambrosio:1998wu} to explain the zenith dependence of the atmospheric neutrino deficit. The average neutrino energy in the CNGS beam was designed to be about 17 GeV, the $\bar{\nu}_{\mu}$ fraction was 2.1\% in terms of expected charged current (CC) interactions, the sum of the $\nu_{e}$ and $\bar{\nu}_{e}$ fractions was below 1\%, while the prompt $\nu_{\tau}$ component was negligible.

The OPERA detector \cite{Acquafredda:2009zz} was located in the underground Gran Sasso Laboratory (LNGS), about 730 km away from the neutrino source at CERN. The detector was a hybrid apparatus made of a nuclear emulsion/lead target complemented by electronic detectors (ED) \cite{OPERA_ed}. The target had a total mass of about 1.25 kt and was composed of two identical sections. Each section consisted of 31 walls of Emulsion Cloud Chamber (ECC) bricks, interleaved by planes of horizontal and vertical scintillator strips (Target Tracker) used to select \cite{Gornushkin:2015eqa} ECC bricks in which a neutrino interaction had occurred. Each ECC brick consisted of 57 emulsion films interleaved with 1 mm thick lead plates, with a $(12.7 \times 10.2)\ \si{cm^2}$ cross section and a total thickness corresponding to about ten radiation lengths. A magnetic muon spectrometer \cite{Ambrosio:2004tm, Acquafredda:2009zz} was positioned downstream of each target section, and was instrumented by resistive plate chambers and high-resolution drift tubes.

The experiment collected data from the CNGS beam from 2008 to 2012, with an integrated exposure of \num{17.97e19} protons on target. A total of 19505 neutrino interaction events in the target were recorded by the electronic detectors, of which 5603 were fully reconstructed in the OPERA emulsion films.

In 2015, the OPERA Collaboration reported the discovery of tau neutrino appearance~\cite{Agafonova:2015jxn}. Five $\nu_{\tau}$ candidate events~\cite{Agafonova:2010dc, Agafonova:2013dtp, Agafonova:2014bcr, Agafonova:2014ptn, Agafonova:2015jxn} were observed, with an expected signal-to-background ratio of $\sim$10. The analysis of these data lead to the exclusion of the background-only hypothesis with a significance of \SI{5.1}{\sigma}.

The final result on $\nu_\tau$ appearance was published in 2018~\cite{Agafonova:2018auq}, using an analysis based on a multivariate approach and a looser selection with respect to \cite{Agafonova:2015jxn}, applied to the complete data set. The achieved significance for tau appearance was $\SI{6.1}{\sigma}$. Additionally, the atmospheric neutrino mass splitting was measured to be $\left|\Delta m^2_{32}\right| = 2.7^{+0.7}_{-0.6} \times 10^{-3} \textrm{eV}^{2}$, assuming maximal two-flavor mixing.

A search for electron neutrino CC interactions was performed \cite{Agafonova:2013xsk, Agafonova:2018dkb}. Unlike $\nu_\tau$ appearance, the sensitivity to the $\nu_\mu \rightarrow \nu_e$ oscillation channel is limited mainly by the $\nu_e$ beam contamination and the systematic uncertainty in the neutrino flux. The collected sample of electron neutrino candidates was in agreement with the expectation from the three-flavor neutrino mixing model and was exploited to obtain a 90\% C.L. upper limit on the parameter $\sin^2 2\theta_{13} < 0.43$ \cite{Agafonova:2018dkb}.

The data were also used to test the hypothesis of one additional light sterile neutrino state, in the $3 + 1$ neutrino mixing model \cite{Agafonova:2013xsk,Agafonova:2015neo,Agafonova:2018dkb}. OPERA is sensitive to a significant part of the allowed region of the sterile neutrino parameters defined by the LSND \cite{Aguilar:2001ty} and MiniBooNE \cite{Aguilar-Arevalo:2018gpe} experiments, and to a large part of the yet unexplored neutrino parameter space related to $\nu_{\tau}$-$\nu_{s}$ mixing.

In this Letter, the final samples of tau and electron neutrino CC interactions are used for the first time in a combined analysis to constrain standard oscillation parameters $\theta_{13}$, $\theta_{23}$, as well as the $3+1$ mixing model ones. In addition, a search for the $\nu_\mu$ disappearance signal was performed using only electronic detector data, assuming three-flavor neutrino mixing.

\section{Data selection}
\paragraph*{Selection of $\nu_\tau$ and $\nu_e$ CC candidate events using emulsion detector data}
The emulsion detector data set consists of events in which the neutrino interaction vertex is fully reconstructed in the emulsion films. Tracking capabilities of the OPERA emulsion technique allow isolating samples of tau and electron neutrino CC interactions with a very high signal-to-background ratio \cite{Agafonova:2013dtp, Agafonova:2014khd}. In the final analysis, ten tau neutrino candidates have been identified \cite{Agafonova:2018auq}, while the expected background amounts to $2.0 \pm 0.4$ events. The sources of background are: \textit{(i)} the decay of charmed particles produced in $\nu_{\mu}$ CC interactions, \textit{(ii)} the reinteraction of hadrons from $\nu_{\mu}$ events in lead, and \textit{(iii)} the large-angle scattering (LAS) of muons produced in $\nu_{\mu}$ CC interactions. Processes \textit{(i)} and $\nu_{\mu}$ CC in \textit{(ii)} represent a background source when the $\mu^{-}$ at the primary vertex is not identified.

The search for electron neutrino appearance led to the selection of 35 $\nu_e$ candidates in the full dataset resulting from a dedicated analysis \cite{Agafonova:2018dkb}. The two main sources of background are: \textit{(i)} $\nu_{\tau}$ CC interactions followed by a $\tau \rightarrow e$ decay and \textit{(ii)} muon-less events with a $\pi^{0}\rightarrow \gamma \gamma$ decay and an $e^{+}e^{-}$ pair from prompt $\gamma$ conversion misidentified as an electron. The total expected background amounts to $1.2 \pm 0.5$ events. Given the prompt $\nu_e$ and $\overline{\nu}_e$ beam components, the number of observed $\nu_{e}$ events is consistent both with no-oscillation (31.9 $\pm$ 3.3) and standard oscillation (34.3 $\pm$ 3.5) hypotheses with normal neutrino mass hierarchy (NH) \cite{Agafonova:2018dkb}.

\paragraph*{Selection of $\nu_\mu$ CC and $\nu$ NC candidate events using only electronic detector data} 
Only electronic detector data were used in a search for $\nu_\mu$ disappearance, since events without a decay topology (most of $\nu_\mu$ CC interactions and $\nu$ NC interactions) are not fully reconstructed in the emulsion target \cite{Agafonova:2014khd}. This analysis is complementary to the appearance study, as this dedicated selection of ED data differs from the one used to locate ECC bricks in which $\nu$ interactions occurred.

The compositions of CC-like and NC-like samples described below were estimated using standard Monte Carlo simulation validated on data \cite{OPERA_ed}, with GENIE v2.8.6 \cite{Andreopoulos:2009rq,Andreopoulos:2015wxa} as a neutrino interaction generator.

A global selection was applied to ED data: \textit{(i)} the interaction vertex must be reconstructed inside the active volume defined in \cite{OPERA_OpCarac}, \textit{(ii)} a fiducial volume cut rejecting events in which neutrinos most probably interacted in the 10 brick walls immediately downstream of a spectrometer, and \textit{(iii)} a cut on the total signal recorded by electronic detectors in order to reject soft non-beam events.

The $\nu_\mu$ CC interaction candidates (CC-like sample) are selected by requiring a muon track with negative charge reconstructed in at least two arms of a single magnetic spectrometer. According to the Monte Carlo simulation, the expected purity of the $\nu_\mu$ CC-like event sample is 99.5\%, and it contains \SI{46}{\%} of expected $\nu_\mu$ CC interactions in the target.

The neutrino NC interaction candidates (NC-like sample) are selected by requiring no track identified as a muon. Assuming global best-fit values of the standard oscillation parameters for NH (Table 14.1 in \cite{Tanabashi:2018oca}), the MC expectation is that 75.3\% of events are $\nu$ and $\overline{\nu}$ NC interactions, 20.2\% are $\nu_\mu$ and $\overline{\nu}_\mu$ CC interactions in which no muon was identified, and 4.5\% are CC interactions of other neutrino flavors. The latter are expected to be: \textit{(i)} prompt $\nu_e$ and $\overline{\nu}_e$ beam components, and \textit{(ii)} flavors appearing through neutrino oscillations, most significantly in channels $\nu_\mu \rightarrow \nu_\tau$ and $\nu_\mu \rightarrow \nu_e$. This sample contains \SI{34}{\%} of expected NC interactions in the target.

NC-like ad CC-like data samples consist of 1724 and 5782 events, respectively.

It should be noted that the estimated composition of NC-like and CC-like samples quoted above do not explicitly enter the analysis, as the appropriate oscillation probability is applied to all MC events as described in the next section.


\section{Analysis}
\paragraph*{Appearance analyses} 
Visible neutrino energy as defined in \cite{Agafonova:2018auq} and reconstructed neutrino energy as defined in \cite{Agafonova:2018dkb} are used as observables for tau and electron neutrino samples, respectively. Their distributions are jointly exploited to test neutrino oscillation phenomena both in standard three-flavor and 3+1 mixing models.

The expected distributions are evaluated using GLoBES \cite{Huber:2004ka,Huber:2007ji}. The smearing matrices, incorporating the detection and selection efficiencies, are calculated by means of MC simulation of the full analysis chain. The simulated CNGS flux \cite{Ferrari}, default neutrino interaction cross sections evaluated by GENIE v2.8.6 \cite{Andreopoulos:2009rq,Andreopoulos:2015wxa}, and the smearing matrices are used as input. It is worth noting that the oscillations of all CNGS beam components ($\nu_{\mu}$, $\bar{\nu}_{\mu}$, $\nu_{e}$ and $\bar{\nu}_{e}$) are taken into account.

The systematic uncertainty on the expected number of tau neutrino candidates is largely dominated by the limited knowledge of the $\nu_{\tau}$ CC interaction cross section and detection efficiency, overall estimated to be 20\% \cite{Agafonova:2018auq}.

The expected number of electron neutrino candidates is affected by systematic uncertainties on: \textit{(i)} the prompt flavor beam components, \textit{(ii)} the $\nu_{e}$ and $\bar{\nu}_{e}$ CC interaction cross sections, and \textit{(iii)} the neutrino detection efficiency. The systematic error on the expected number of electron neutrino candidates is estimated to be 20\% for reconstructed energies below 10 GeV and 10\% otherwise \cite{Agafonova:2018dkb}.

The statistical analysis of the data is based on a maximum-likelihood joint fit across the two samples. The data are binned and the overall likelihood ($\mathcal{L}$) is constructed as the product of the contribution for each data sample ($\mathcal{L}_{\tau}$, $\mathcal{L}_{e}$) and a Gaussian constraint $\mathcal{G}$ on $\Delta m^{2}_{31}$ as:
\begin{equation}\label{eq:appearance_likelihood}
  \mathcal{L} = \mathcal{L}_{e} \times \mathcal{L}_{\tau} \times \mathcal{G} \left( \overline{\Delta m^{2}_{31}}|\mu_{31},\overline{\sigma_{31}} \right)\;,
\end{equation}
where $\overline{\Delta m^{2}_{31}}$ and $\overline{\sigma_{31}}$ are the best fit value and the $\SI{1}{\sigma}$ uncertainty from a global fit of the current neutrino oscillation data (Table 14.1 in \cite{Tanabashi:2018oca}). The used values for $\overline{\Delta m^{2}_{31}} \pm  \overline{\sigma_{31}}$ are $\left(2.56 \pm 0.05 \right) \times 10^{-3} \textrm{ eV}^{2}$ and $\left(-2.49 \pm 0.05 \right) \times 10^{-3} \textrm{ eV}^{2}$ for NH and inverted hierarchy (IH), respectively. The floating parameter $\mu_{31}$ represents the true value of the squared mass difference. 

The negative log-likelihood functions $- \ln \mathcal{L}_{x}$ are defined as:
\begin{eqnarray}\label{eq:likelihood}
  - \ln \mathcal{L}_{x} &=& \sum_{i=1}^{N^{(x)}} \left(\mu_{i}^{(x)} - n_{i}^{(x)} \ln \mu_{i}^{(x)}\right) \nonumber \\
  &+& \frac{1}{2}\sum^{n_{sys}^{(x)}}_{j=1} \left(\frac{\phi^{(x)}_{j}-\hat{\phi}^{(x)}_{j}}{\sigma^{(x)}_{\phi_{j}}}\right)^{2}\;,
\end{eqnarray}
with $x = \tau \textrm{ or } e$, $N^{(x)}$ is the number of bins, $n_{i}$ is the number of events in the \textit{i}-th bin of the data histogram, and $\mu_{i} = \mu_{i}\left(\bar{\theta},\bar{\phi}\right)$ is the expected number of events in the $i$-th bin from the set of oscillations parameters $\bar{\theta}$, namely the mixing matrix elements $U_{\alpha\beta}$ and the squared mass splittings $\Delta m^{2}_{ij}$, and the set of parameters $\bar{\phi}$ which account for systematic uncertainties. The second term of Eq. \ref{eq:likelihood} accounts for the a priori knowledge of the systematic uncertainties, where $\hat{\phi}_{j}$ and $\sigma_{\phi_{j}}$ are respectively the estimated value and the uncertainty on parameter $\phi_{j}$. The non-zero value of $\Delta m^{2}_{21}$ is taken into account as well as matter effects assuming a constant Earth crust density estimated with the PREM \cite{Dziewonski:1981xy,Stacey} onion shell model, even though they were checked to be negligible for $\Delta m^2_{41} > 1 \textrm{ eV}^{2}$.

\paragraph*{Disappearance analysis} A search for the $\nu_\mu$ disappearance signal is performed in the dedicated ED data sample. Since the experiment was not equipped with a near detector, the uncertainty on the $\nu_{\mu}$ flux is estimated to be \SI{15}{\%}. On the other hand, the detector was located far from the first oscillation maximum, so that the $\nu_\mu$ CC rate reduction at the detector due to oscillation does not exceed 2\% in the standard 3-flavor model. To overcome these difficulties, a NC-like/CC-like event rate ratio is used as an estimator of the disappearance oscillation probability that is independent from the overall flux normalization uncertainty. 

Since it is experimentally unfeasible to reconstruct the energy of an incoming neutrino in NC events, NC-like/CC-like event rate is evaluated as a function of a measurement of the energy deposited in the target, denoted by $E_\text{T}$.


In order to constrain the value of the atmospheric mass splitting $\Delta m^2_{32}$, three histograms with identical binning in $E_\text{T}$ are constructed: \textit{(i)} the predicted NC-like/CC-like ratio evaluated using MC simulation as a function of $\Delta m^2_{32}$, \textit{(ii)} the observed number of CC-like events, \textit{(iii)} the observed number of NC-like events.

While constructing histogram \textit{(i)}, global best-fit values (Table 14.1 in \cite{Tanabashi:2018oca}) for all oscillation parameters except the parameter of interest $\Delta m^2_{32}$ are used, separately for NH and IH. For each value of $\Delta m^2_{32}$, a proper oscillation probability is calculated for each simulated event in both NC-like and CC-like samples, including disappearance of $\nu_\mu$, $\overline{\nu}_\mu$, $\nu_e$ and $\overline{\nu}_e$ CC interactions and $\nu_\mu \rightarrow \nu_\tau$ and $\nu_\mu \rightarrow \nu_e$ appearance.

A dedicated statistical model is built to properly construct confidence intervals based on a ratio of two Poisson random variables. Given two outcomes $l$ and $k$ of two Poisson distributions with mean values $\lambda_l$ and $\lambda_k$, respectively, the Bayesian probability of ratio $x \equiv \lambda_l / \lambda_k$, assuming a flat prior on $x$, is provided by the formula
\begin{equation}
\label{eq:ana_bayes_ratio}
P_{l/k}(x) = \frac{(k+l+1)!}{k!\ l!} \frac{x^l}{(1+x)^{k+l+2}}\;.
\end{equation}

The likelihood function is constructed as
\begin{equation}
L(\Delta m^2_{32}) = \prod_{j=1}^N P_{\text{NC}_j / \text{CC}_j}\left( R_j (\Delta m^2_{32}) \right)\;,
\label{eq:ana_likelihood_used}
\end{equation}
where $j$ is a bin label running on all three histograms; $\text{NC}_j$ and $\text{CC}_j$ are the $j$-th bin content of the NC-like \textit{(ii)} and CC-like \textit{(iii)} histograms, respectively; $R_j (\Delta m^2_{32})$ is the NC-like/CC-like ratio obtained from histogram \textit{(i)} as a function of oscillation parameters; $P_{\text{NC}_j / \text{CC}_j}\left( x \right)$ is a probability distribution defined in Eq.~\ref{eq:ana_bayes_ratio}.

In order to extract the confidence interval, the upper limit test statistic \cite{Cowan:2010js} is used, defined by equation
\begin{equation}
\label{eq:ana_q_mu_dm32}
q_{\Delta m_{32}^2} = 
\begin{cases}
-2 \ln \frac{L(\Delta m^2_{32})}{L(\widehat{\Delta m_{32}^2})} & \widehat{\Delta m_{32}^2} \leq \Delta m^2_{32} \\
0 &  \widehat{\Delta m_{32}^2} > \Delta m_{32}^2 
\end{cases}\;,
\end{equation}
where $\widehat{\Delta m_{32}^2}$ is the value of $\Delta m_{32}^2$ which maximizes the likelihood function $L(\Delta m^2_{32})$. A complete treatment of the statistical model can be found in \cite{Budimir_PHD:2018}.

The effect of uncorrelated uncertainties in the knowledge of the CNGS flux is estimated by dividing the simulated neutrino events into subsets (bins) according to true neutrino energy. The bins are \SI{10}{GeV} wide, and the weight of every event in a single bin is multiplied by the same Gaussian random number with mean value of $1$ and standard deviation of $0.15$, corresponding to the uncertainty on the total neutrino flux. Event weights in different bins are multiplied by different random numbers from the same Gaussian distribution. A set of $1000$ different smeared fluxes is generated in this way for each value of $\Delta m^2_{32}$.

The likelihood function is modified to take into account these multiple smeared fluxes as follows:
\begin{equation}\label{eq:ana_smeared_likelihood}
L(\Delta m^2_{32}) = \prod_{i=1}^N \left( \frac{1}{M} \sum_{j=1}^M P_{\text{NC}_i / \text{CC}_i}\left( R_i^{(j)} (\Delta m^2_{32}) \right) \right)\;,
\end{equation}
where $M$ is the number of smeared fluxes, and $R_i^{(j)} (\Delta m^2_{32})$ is a MC predicted NC-like over CC-like event ratio in the $i$-th $E_\text{T}$ bin obtained using $j$-th smeared input flux.

\section{Results}
\begin{figure}[!ht] 
\centering
\includegraphics[width=1.0\columnwidth]{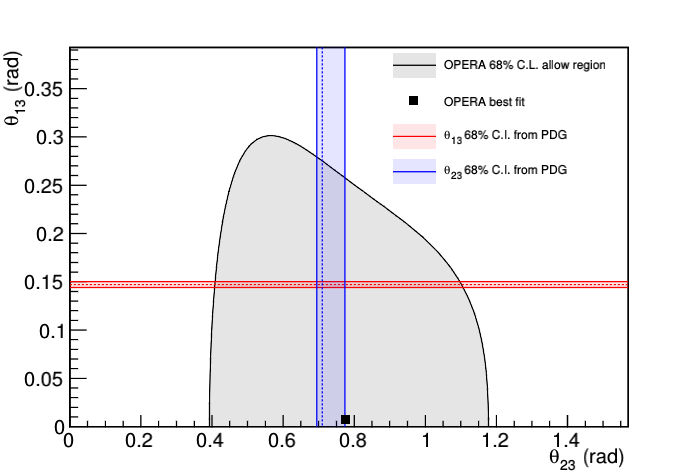}
\caption{OPERA 68\% C.L. allowed region in the $\theta_{13}$ and $\theta_{23}$ parameter space for the normal hierarchy of the three standard neutrino masses. Red and blue dashed lines (areas) represent 1$\sigma$ confidence interval obtained from the global best fit values (Table 14.1 in \cite{Tanabashi:2018oca}, 3$\sigma$ allowed ranges) for $\theta_{13}$ and $\theta_{23}$ respectively.}
\label{fig:theta13_23}
\end{figure}

\paragraph*{Standard three-flavor model} Appearance channels are jointly used to obtain constraints on three-flavor model parameters $\theta_{23}$ and $\theta_{13}$. All other parameters are treated as nuisance and the profile likelihood ratio is used to remove the dependence of the likelihood (Eq.~\ref{eq:appearance_likelihood}) on them. The 68\% C.L. allowed region for NH, obtained using Wilks' theorem, is reported in Fig.~\ref{fig:theta13_23}. Resulting best fit value and $\SI{1}{\sigma}$ confidence interval for $\theta_{23}$ parameter are $\theta_{23} = 0.78^{+0.32}_{-0.31} \textrm{ rad}$; the $\SI{1}{\sigma}$ confidence interval for $\theta_{13}$ is $\left[0,0.20\right] \textrm{rad}$. The best fit value of the constrained parameter $\Delta m^{2}_{31}$ is $2.56 \times 10^{-2}$ eV$^{2}$.

\begin{figure}[!ht] 
\centering
\includegraphics[width=\columnwidth]{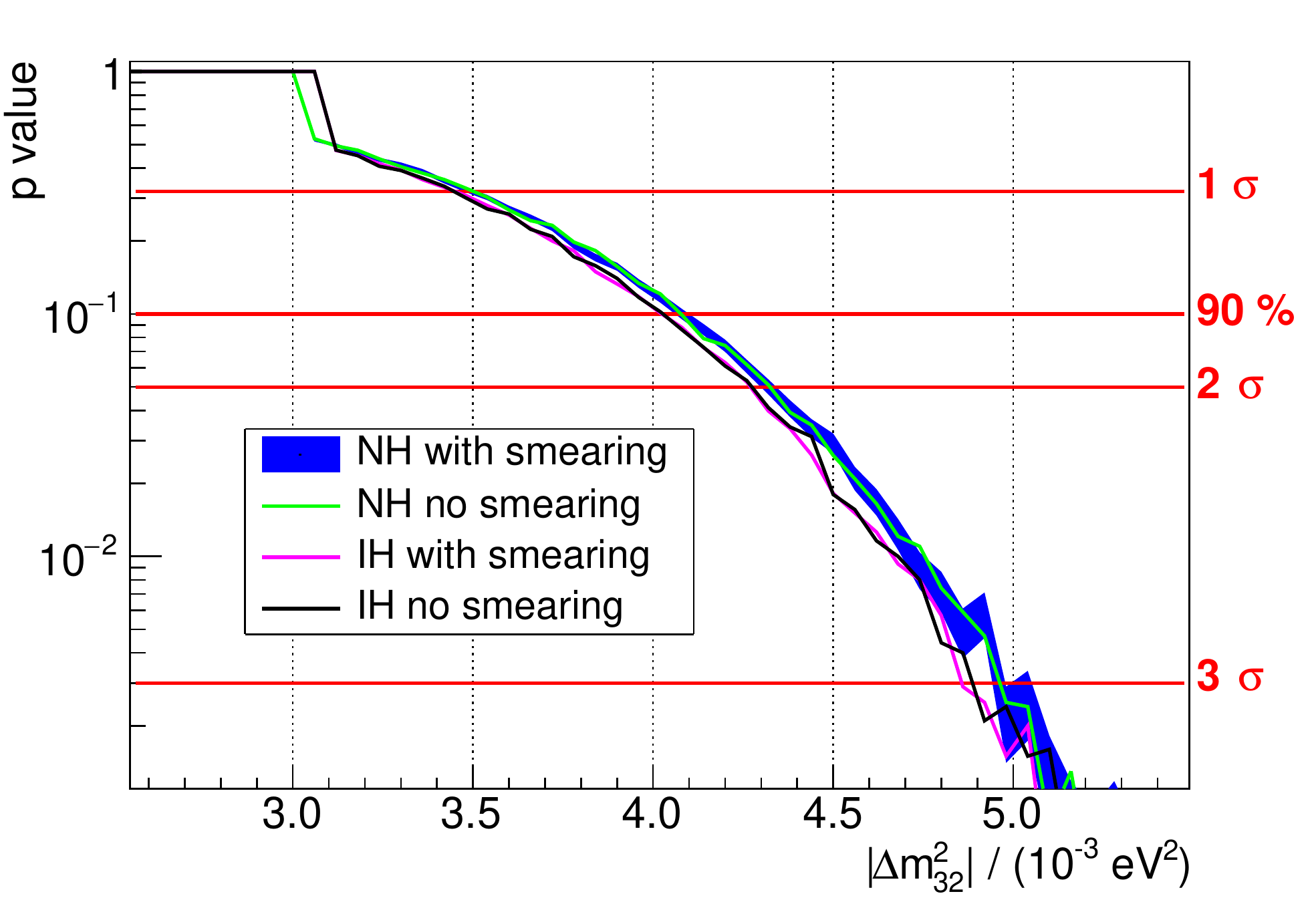}
\caption{$p$-values as a function of $\abs{\Delta m^2_{32}}$, with all other oscillation parameters fixed to central values (Table 14.1 in \cite{Tanabashi:2018oca}). Vertical width of the \texttt{"NH with smearing"} curve represents the \SI{90}{\percent} C.L. interval on p-value due to the finite numbers of pseudo-experiments; other curves have similar uncertainty which is not shown for clarity. $p$-values for $\abs{\Delta m^2_{32}} < \SI{2.5e-3}{eV^2}$ are equal to 1 in all cases.}
\label{fig:numu_disappearance}
\end{figure}

The $\nu_\mu \rightarrow \nu_\mu$ disappearance channel was explored using the test statistic (Eq.~\ref{eq:ana_q_mu_dm32}) based on the likelihood (Eq.~\ref{eq:ana_likelihood_used}) to obtain an upper limit on $\Delta m_{32}^2$. The values of $\Delta m_{32}^2$ are sampled on a grid in the interval $\left[ 0.0, 6.0 \right] \times 10^{-3}\text{ eV}^2$ for NH, and $\left[ -6.0, 0.0 \right] \times 10^{-3}\text{ eV}^2$ for IH. In both cases, the grid spacing is $6.0 \times 10^{-5}\text{ eV}^2$. The distribution of $q_{\Delta m_{32}^2}$ (Eq.~\ref{eq:ana_q_mu_dm32}) is calculated for each of these grid points using 10 000 pseudo-experiments. This is then used to calculate a $p$-value for each point for both NH and IH cases, shown in Fig.~\ref{fig:numu_disappearance} as \verb!"no smearing"! curves. For both NH and IH assumptions, the \SI{90}{\percent} confidence interval is given as $\left| \Delta m^2_{32}  \right| < \SI{4.1e-3}{eV^2}$.

The p-values obtained using smeared neutrino fluxes by using likelihood (Eq.~\ref{eq:ana_smeared_likelihood}) are shown in Fig.~\ref{fig:numu_disappearance} as \verb!"with smearing"! curves. There is no significant difference w.r.t $p$-values obtained without smearing.

\paragraph*{Sterile 3+1 model}
The excess of electron neutrino and antineutrino events reported by the LSND \cite{Aguilar:2001ty} and MiniBooNE \cite{Aguilar-Arevalo:2018gpe} experiments may be interpreted as due to the presence of light $\mathcal{O}(1 \textrm{ eV})$ sterile neutrinos. OPERA can test this hypothesis searching for deviation from the prediction of the three-flavor neutrino model. The oscillation probability in this model depends on a four by four unitary mixing matrix $U$ and three squared mass differences ($\Delta m^2_{21}$, $\Delta m^{2}_{31}$ and $\Delta m^{2}_{41}$) \cite{Dentler:2018sju}.

Here, for the first time, tau and electron neutrino appearance channels are jointly used to test the sterile neutrino hypothesis in the framework of the $3+1$ model. Since the NC-like over CC-like event ratio is poorly sensitive to the effects induced by the presence of a sterile neutrino state, the muon neutrino disappearance channel is not exploited in this analysis.

\begin{figure}[!ht] 
\centering
\includegraphics[width=0.9\columnwidth]{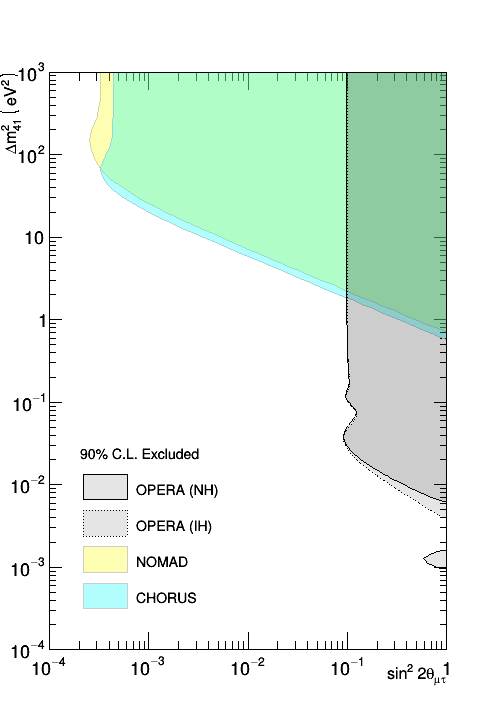}
\caption{OPERA 90\% CL exclusion region in the $\Delta m^{2}_{41}$ and $\sin^{2} 2\theta_{\mu\tau}$ parameter space for the normal (NH, solid line) and inverted (IH, dashed line) hierarchy of the three standard neutrino masses. The exclusion regions by NOMAD \cite{Astier:2001yj} and CHORUS \cite{Eskut:2007rn} are also shown.}
\label{fig:mutau}
\end{figure}

\begin{figure}[!ht] 
\centering
\includegraphics[width=0.9\columnwidth]{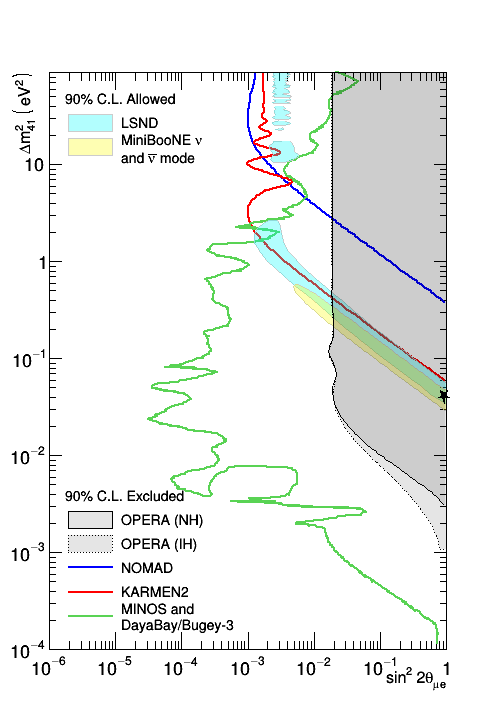}
\caption{The 90\% C.L. exclusion region in the $\Delta m_{41}^{2}$ and $\sin^{2}2\theta_{\mu e}$ plane is shown for the normal (NH, solid line) and inverted (IH, dashed line) hierarchy of the three standard neutrino masses. The plot also reports the 90\% C.L. allowed region obtained by LSND \cite{Aguilar:2001ty} (cyan) and MiniBooNE combining $\nu$ and $\bar{\nu}$ mode \cite{Aguilar-Arevalo:2018gpe} (yellow). The blue and red lines represent the 90\% C.L. exclusion regions obtained in appearance mode by NOMAD \cite{Astier:2003gs} and KARMEN2 \cite{Armbruster:2002mp}, respectively. The 90\% C.L. exclusion region obtained in disappearance mode by the MINOS and DayaBay/Bugey-3 joint analysis \cite{Adamson:2016jku} is shown as green line. The black star ($\star$) corresponds to the MiniBooNE best-fit values for the combined analysis of $\nu$ and $\bar{\nu}$ data.}
\label{fig:mue}
\end{figure}

Defining $\sin^{2}2\theta_{\mu\tau} = 4|U_{\tau4}|^2|U_{\mu4}|^2$, $\sin^{2}2\theta_{\mu e} = 4|U_{e4}|^2|U_{\mu4}|^2$ and $\Delta m^{2}_{41}$ as the parameters of interest, the profile likelihood ratio is used to remove the dependence of the likelihood (Eq. \ref{eq:appearance_likelihood}) on the other parameters treated as nuisance. Exclusion regions of $\Delta m^{2}_{41}$ versus $\sin^22\theta_{\mu\tau}$ and $\sin^22\theta_{\mu e}$ are shown in Fig. \ref{fig:mutau} and Fig. \ref{fig:mue}, respectively. The result is restricted to positive $\Delta m^{2}_{41}$ values since negative values are disfavored by results on the sum of neutrino masses from cosmological surveys \cite{Ade:2015xua}. The MINOS+ Collaboration has recently reported an analysis setting limits on the existence of a sterile neutrino state \cite{Adamson:2017uda}.

For $\Delta m^{2}_{41} > 0.1 \textrm{ eV}^{2}$, the upper limits on $\sin^{2}2\theta_{\mu\tau}$ and $\sin^{2}2\theta_{\mu e}$ are set to 0.10 and 0.019 both for the case of NH and IH. The limits obtained on the sterile oscillation parameters improve those reported in \cite{Agafonova:2018dkb,Agafonova:2015neo}. The values of the oscillation parameters $\left(\Delta m^{2}_{41} = 0.041 \textrm{ eV}^{2},\, \sin^{2}2\theta_{\mu e} = 0.92\right)$ corresponding to the MiniBooNE combined neutrino and antineutrino best-fit \cite{Aguilar-Arevalo:2018gpe} are excluded with a $p$-value of $8.9\times10^{-4}$, corresponding to a significance of 3.3 $\sigma$.

\section{Conclusions}
Searches for $\nu_\tau$ and $\nu_e$ appearance and for $\nu_\mu$ disappearance were performed using the full OPERA data sample.

The data are compatible with the three-flavor neutrino model and constraints on $\theta_{23}$ and $\theta_{13}$ were derived jointly for the first time exploiting tau and electron neutrino appearance channels. A best fit value of $\theta_{23} = 0.78^{+0.32}_{-0.31} \textrm{ rad}$ at $\SI{1}{\sigma}$ C.L. is obtained, while $\theta_{13}$ is constrained to $\left[0,0.20\right] \textrm{rad}$ at $\SI{1}{\sigma}$ C.L.

Additionally, a dedicated sample of OPERA electronic detector data is used to perform a search for the $\nu_\mu$ disappearance signal in the CNGS beam. Assuming all other mixing parameters equal to the global fit central values (Table 14.1 in \cite{Tanabashi:2018oca}), an upper limit $\left| \Delta m^2_{32}  \right| < \SI{4.1e-3}{eV^2}$ at $\text{90\% C.L.}$ is obtained.

Finally, $\nu_\tau$ and $\nu_e$ appearance channels were combined for the first time to constrain parameters of the $3+1$ sterile mixing model. For $\Delta m^{2}_{41} > 0.1 \textrm{ eV}^{2}$, upper limits on $\sin^{2}2\theta_{\mu\tau}$ and $\sin^{2}2\theta_{\mu e}$ are set to $0.10$ and $0.019$ for NH and IH. The MiniBooNE best-fit \cite{Aguilar-Arevalo:2018gpe} values $\left(\Delta m^{2}_{41} = 0.041 \textrm{ eV}^{2}, \sin^{2}2\theta_{\mu e} = 0.92\right)$ are excluded with 3.3 $\sigma$ significance.

\section*{Acknowledgements}
\footnotesize{We warmly thank  CERN for the successful operation of the CNGS facility and INFN for the continuous support given by hosting the experiment in its LNGS laboratory. Funding is gratefully acknowledged from  national agencies and Institutions supporting us, namely: Fonds de la Recherche Scientifique-FNRS and Institut Interuniversitaire des Sciences Nucleaires for Belgium; MZO for Croatia; CNRS and IN2P3 for France; BMBF for Germany; INFN for Italy; JSPS, MEXT, the QFPU-Global COE program of Nagoya University, and Promotion and Mutual Aid Corporation for Private Schools of Japan for Japan; SNF, the University of Bern and ETH Zurich for Switzerland; the Programs of the Presidium of the Russian Academy of Sciences (Neutrino Physics and Experimental and Theoretical Researches of Fundamental Interactions), and the Ministry of Education and Science of the Russian Federation for Russia, the Basic Science Research Program through the National Research Foundation of Korea (NRF) funded by the Ministry of Science and ICT (Grant No. NRF-2018R1A2B2007757) for Korea; and TUBITAK, the Scientific and Technological Research Council of Turkey for Turkey (Grant No. 108T324). We warmly acknowledge the important contributions given by our deceased colleagues G. Giacomelli, A. Ljubičić, G. Romano and P. Tolun.}

\end{document}